\documentclass[reprint,superscriptaddress,showpacs,amsmath,amssymb,prl,aps]{revtex4-1}

\usepackage{amsmath}
\usepackage{amssymb}
\usepackage{graphics}
\usepackage{graphicx}
\usepackage{dcolumn}
\usepackage{txfonts}
\usepackage[pdfpagemode=UseNone,pdfstartview=FitH,colorlinks=true,linkcolor=blue,urlcolor=blue,anchorcolor=blue,citecolor=blue]{hyperref}
\allowdisplaybreaks[4]

\begin{document}
\title{Bistability of Cavity Magnon Polaritons}
\author{Yi-Pu Wang}
\affiliation{Quantum Physics and Quantum Information Division, Beijing Computational Science Research Center, Beijing 100193, China}
\author{Guo-Qiang Zhang}
\affiliation{Quantum Physics and Quantum Information Division, Beijing Computational Science Research Center, Beijing 100193, China}
\author{Dengke Zhang}
\thanks{Present address:~Department of Engineering, University of Cambridge, Cambridge CB3 0FA, United Kingdom}
\affiliation{Quantum Physics and Quantum Information Division, Beijing Computational Science Research Center, Beijing 100193, China}
\author{Tie-Fu Li}
\thanks{litf@tsinghua.edu.cn}
\affiliation{Institute of Microelectronics, Tsinghua National Laboratory of Information Science and Technology, Tsinghua University, Beijing 100084, China}
\affiliation{Quantum Physics and Quantum Information Division, Beijing Computational Science Research Center, Beijing 100193, China}
\author{C.-M. Hu}
\affiliation{Department of Physics and Astronomy, University of Manitoba, Winnipeg R3T 2N2, Canada}
\author{J. Q. You}
\thanks{jqyou@csrc.ac.cn}
\affiliation{Quantum Physics and Quantum Information Division, Beijing Computational Science Research Center, Beijing 100193, China}
\date{\today}
\begin{abstract}
We report the first observation of the magnon-polariton bistability in a cavity magnonics system consisting of cavity photons strongly interacting with the magnons in a small yttrium iron garnet (YIG) sphere. The bistable behaviors are emerged as sharp frequency switchings of the cavity magnon-polaritons (CMPs) and related to the transition between states with large and small number of polaritons. In our experiment, we align, respectively, the [100] and [110] crystallographic axes of the YIG sphere parallel to the static magnetic field and find very different bistable behaviors (e.g., clockwise and counter-clockwise hysteresis loops) in these two cases. The experimental results are well fitted and explained as being due to the Kerr nonlinearity with either positive or negative coefficient. Moreover, when the magnetic field is tuned away from the anticrossing point of CMPs, we observe simultaneous bistability of both magnons and cavity photons by applying a drive field on the lower branch.
\end{abstract}
\pacs{75.30.Ds, 71.36.+c, 42.65.Pc}
\maketitle

Both quantum information processing~\cite{Wallquist-09} and future quantum internet~\cite{Kimble-08} inevitably need efficient quantum information transfers among different physical systems. Hybrid quantum systems may provide hopeful solutions to these problems~\cite{Xiang-13,Kurizki-15}. Recently, cavity magnonics has attracted much interest (see, e.g., \cite{Huebl-13,Tabuchi-14,Zhang-14,Tobar-14,Hu-15,You-15,Soykal,Bauer-15}), which involves cavity photons strongly or ultrastrongly~\cite{Bourhill-16} interacting with collective spin excitations in a millimeter-scale yttrium iron garnet (YIG) crystal. This hybrid system gives rise to new quasiparticles called cavity magnon-polaritons (CMPs)~\cite{Cao-15,Hu-15-2}. Benefiting from low damping rates of both magnons and cavity photons, this hybrid system is also expected to become a building block of the future quantum information network. Now, a versatile quantum information processing platform based on the coherent couplings among magnons, cavity microwave photons~\cite{Huebl-13,Tabuchi-14,Zhang-14}, optical photons~\cite{Osada-16,Zhangxu-16,haighprl-16,Braggio-17}, phonons~\cite{zhangxuSA-16} and superconducting qubits~\cite{TabuchiScience-15,NakamuraSA-17} is being established, with the strong coupling between magnons and cavity photons being the core of the hybrid quantum system.

For the cavity magnonics system, in addition to the hybridization between magnons and cavity photons, nonlinear effect can also play an important role. Originating from the magnetocrystalline anisotropy in the YIG~\cite{Stancil-2009,Gurevich-1996}, this nonlinearity is related to the Kerr effect of magnons and the nonlinearity-induced frequency shift has been demonstrated in the dispersive regime at cryogenic temperature~\cite{Wang-16}. In this Letter, we observe the CMP bistability in a cavity magnonics system.
In a wide range of parameters, the frequency of the CMPs is found to jump up or down sharply at the switching points where the CMPs transition from one state to another. To clearly demonstrate the nonlinear effect, we implement the experiment by aligning the [100] and [110] crystallographic axes of the YIG sphere parallel to the externally applied static magnetic field, respectively. The measured results show very different features in these two cases, such as the blue- or red-shift of the frequency of the CMPs and the emergence of the clockwise or counter-clockwise hysteresis loop related to the CMP bistability.
Also, we theoretically fit the experimental results well and explain the very different phenomena of the CMPs as being due to the Kerr nonlinearity with either a {\it positive} or {\it negative} coefficient.

To our knowledge, this work is the first convincing observation of the bistability in CMPs. As an important nonlinear phenomenon, bistability is not only of fundamental interest in studying dissipative quantum systems~\cite{Rodriguez-17,Letscher-17}, but also has potential applications in switches~\cite{Paraiso-10,Bilal-17} and memories~\cite{Kuramochi-14,Kubytskyi-14}.
Our cavity magnonics system offers a new platform to explore these applications. Also, the employed Kittle-mode magnons have distinct merits, such as the tunability with magnetic field and the maintenance of good quantum coherence even at room temperature~\cite{Zhang-14,You-15}. These advantages may bring new possibilities in exploring nonlinear properties of the system. Indeed, we also observe the simultaneous bistability of both magnons and cavity photons
at a very off-resonance point by applying a drive field on the lower branch, where the optical bistability is achieved via the magnetic bistability. This observation shows that the CMPs can serve as a bridge/transducer between optical and magnetic bistabilities, which paves a new way for using one effect to induce and control the other.

\begin{figure}[htbp]
  \centering
  \includegraphics[width=0.48\textwidth]{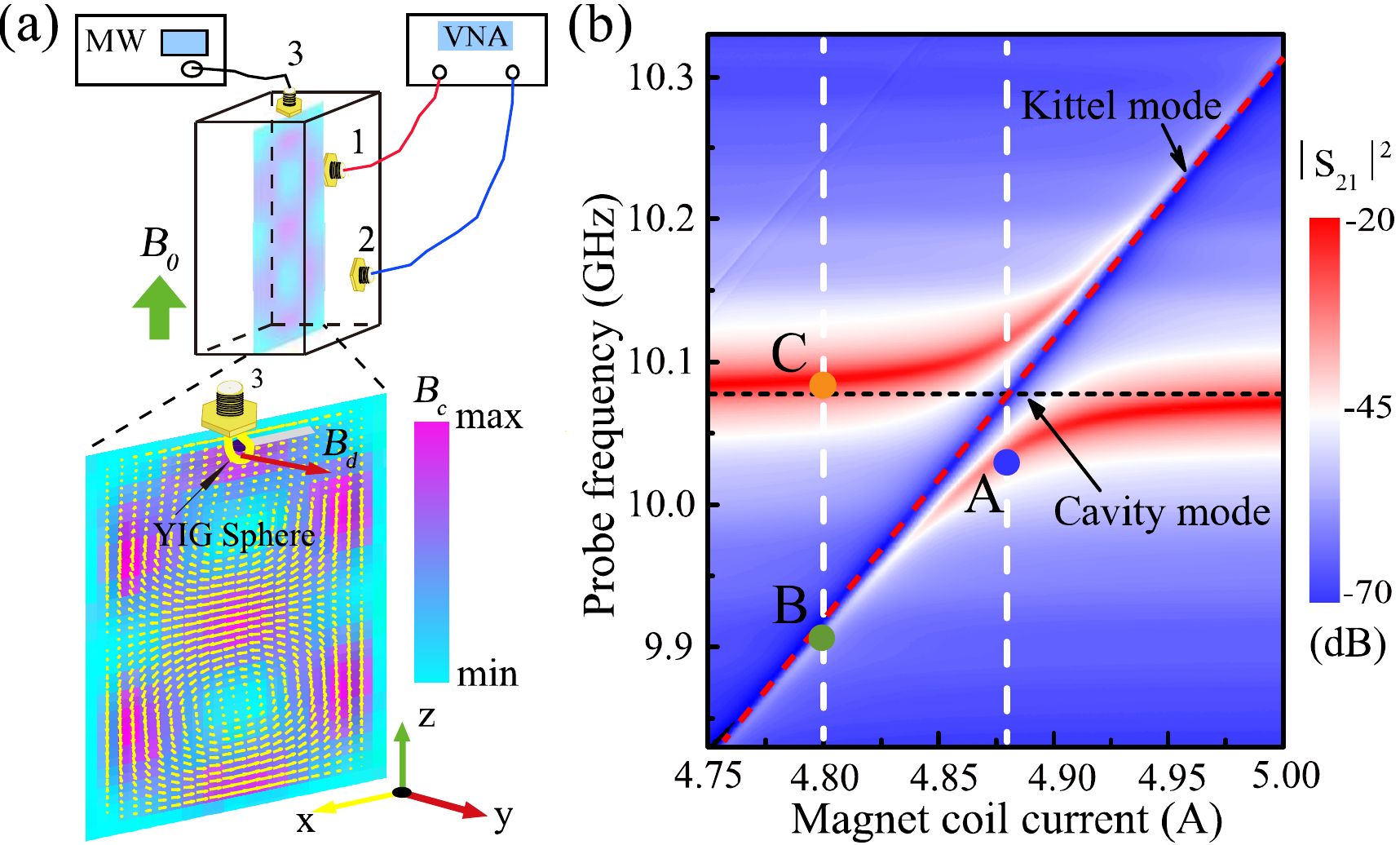}
  \caption{(a)~Schematic diagram of the experimental setup. The 3D cavity with a small YIG sphere embedded is placed in the static magnetic field $B_0$ generated by an electromagnet. Ports 1 and 2 of the cavity, connected to a vector network analyzer (VNA), are used for transmission spectroscopy and port 3 connected to a microwave (MW) source is for driving the YIG sphere. At the bottom, the magnetic-field distribution of the $\rm{TE}_{\rm{102}}$ mode is magnified for clarity. The small YIG sphere is placed beside a circular-loop antenna and located at the magnetic-field antinode of the $\rm{TE}_{\rm{102}}$ mode. (b)~Transmission spectrum of the CMPs measured versus the magnet coil current (i.e., the static magnetic field) and the frequency of the probe field. Two vertical dashed lines indicate, respectively, the resonance and the very off-resonance points at which we show the CMP bistability.}
  \label{fig:1}
\end{figure}

The experimental setup for our hybrid system is schematically shown in Fig.~\ref{fig:1}(a). We use a three-dimensional (3D) rectangular cavity made of oxygen-free copper with inner dimensions $44.0\times22.0\times6.0~\rm{mm^3}$ and have a small YIG sphere of diameter 1~mm glued on an inner wall of the cavity at a magnetic-field antinode of the cavity mode $\rm{TE}_{\rm{102}}$. The cavity has three ports, with ports 1 and 2 as the input and output ports for measuring transmission spectrum and also with a specially designed port (i.e., port 3) in the vicinity of the YIG sphere for conveniently loading a microwave  drive field via a loop antenna. Here we focus on the Kittel mode which is a spatially uniform mode of the ferromagnetic spin waves~\cite{Kittel-1948}. To reduce the disturbance from other magnetostatic modes~\cite{Walker-1958,Bell-1959}, the small YIG sphere is placed at the uniform field of the antenna. The whole cavity with the YIG sphere embedded is placed in a static magnetic field $B_0$ created by a high-precision tunable electromagnet at room temperature. This bias magnetic field, the magnetic component of the microwave drive field, and the magnetic field of the $\rm{TE}_{\rm{102}}$ mode are nearly perpendicular to one another at the site of the small YIG sphere.

When the frequency of the Kittel-mode magnons is tuned in resonance with the microwave photons of the cavity mode $\rm{TE}_{\rm{102}}$, anticrossing of energy levels occurs owing to the strong coupling between magnons and cavity photons, which gives rise to two branches of CMPs [see Fig.~\ref{fig:1}(b)].
The magnon-photon coupling strength is found to be $g_{\rm{m}}/2\pi=41$~MHz from the energy splitting at the resonance (anticrossing) point. The fitted cavity-mode linewidth is $\kappa/2\pi\equiv(\kappa_{1}+\kappa_{2}+\kappa_{3}+\kappa_{\rm{int}})/2\pi=3.8$~MHz, where $\kappa_{i}$ ($i=1,2,3$) is the decay rate of the cavity due to the $i$th port and $\kappa_{\rm{int}}$ is the intrinsic loss of the cavity. Also, the Kittel-mode linewidth is found to be $\gamma_{\rm{m}}/2\pi=17.5$~MHz. It is clear that the system is in the strong-coupling regime because $g_{\rm{m}}>\kappa,\gamma_{\rm{m}}$.

\begin{figure}
\includegraphics[width=0.48\textwidth]{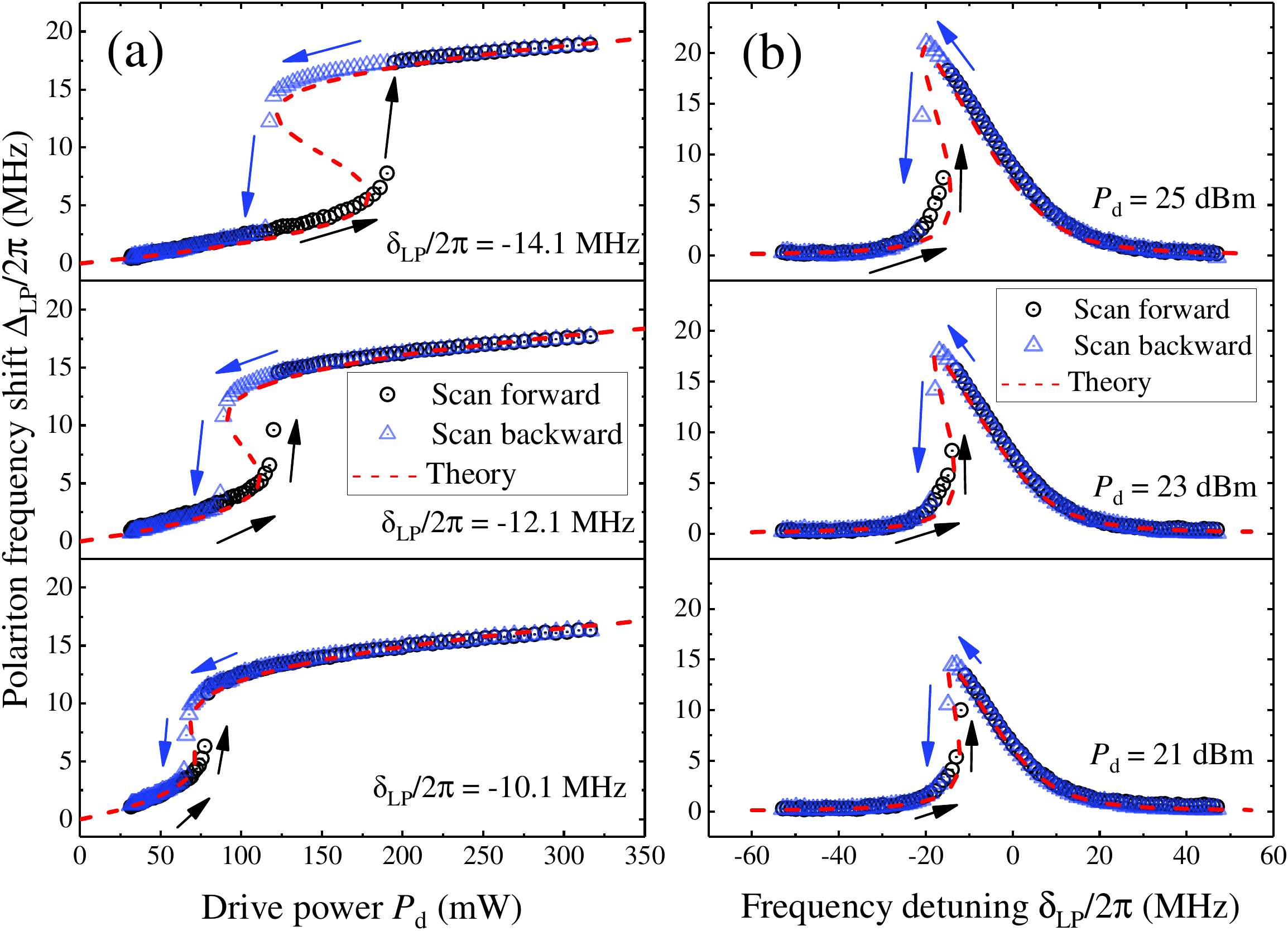}
\caption{(a)~$\Delta_{\rm LP}/2\pi$ versus $P_{\rm d}$ when $\delta_{\rm{LP}}/2\pi=-14.1$, $-12.1$, and $-10.1$~MHz. (b)~$\Delta_{\rm LP}/2\pi$ versus $\delta_{\rm{LP}}/2\pi$ when $P_{\rm d}=25$, 23, and 21~dBm.
The black circle (blue triangle) dots are the forward (backward)-scanning results.
The red dashed curves are theoretical results obtained using Eq.~(\ref{cubic}), with $\gamma_{\rm{LP}}/2\pi=(\gamma_{\rm{m}}+\kappa)/4\pi=10.65$~MHz. In (a), the characteristic constant $c/(2\pi)^3$ is fitted to be $3.15$, $3.55$, and $3.82$~MHz$^{3}$/mW for curves from top to bottom. In (b), $c/(2\pi)^3$ is fitted to be $1.85$, $2.52$, and $3.22$~MHz$^{3}$/mW for curves from top to bottom.
The [100] crystallographic axis of the YIG sphere is aligned parallel to the static magnetic field $B_0$ and the magnons are in resonance with the cavity mode $\rm{TE}_{\rm{102}}$.}
\label{fig:2}
\end{figure}

Here we demonstrate the nonlinear effect in the cavity magnonics system by driving the YIG sphere with a microwave field via the loop antenna. We first focus on the case with the magnons in resonance with the cavity photons, where a CMP is the maximal superposition of a magnon and a cavity photon. Also, the YIG sphere is aligned to have its [100] crystallographic axis parallel to the static magnetic field $B_0$. In Fig.~\ref{fig:2}(a), we measure the frequency shift $\Delta_{\rm LP}$ of the {\it lower}-branch CMPs versus the drive power $P_{\rm d}$ for different values of the drive-field frequency detuning (see \cite{supplementary} for the measurement method). Here the angular frequency $\omega_{\rm{LP}}$ of the {\it lower}-branch CMPs is tuned to be at the anticrossing point $A$ in Fig.~\ref{fig:1}(b) and the drive-field frequency detuning $\delta_{\rm{LP}}\equiv \omega_{\rm{LP}}-\omega_{\rm{d}}$ is relative to the {\it lower}-branch CMPs, where $\omega_{\rm{d}}$ is the angular frequency of the drive field. At the resonance point with $\omega_{\rm m}=\omega_{\rm c}$ (where $\omega_{\rm m}$ is the frequency of the Kittel-mode magnons and $\omega_{\rm c}$ is the frequency of the cavity mode $\rm{TE}_{\rm{102}}$), a hysteresis loop is clearly seen at $\delta_{\rm{LP}}<0$, revealing the emergence of the CMP bistability in the cavity magnonics system. This hysteresis loop is counter-clockwise when considering the increasing and decreasing directions of the drive power. Moreover, its area reduces when decreasing the frequency detuning $|\delta_{\rm{LP}}|$. In Fig.~\ref{fig:2}(b), we measure the frequency shift $\Delta_{\rm LP}$ of the same {\it lower}-branch CMPs versus the frequency detuning $\delta_{\rm{LP}}$ for different values of $P_{\rm d}$. When the increasing and decreasing directions of $\delta_{\rm{LP}}$ are considered, a counter-clockwise hysteresis loop is also clearly shown, and its area decreases when reducing $P_{\rm d}$.

For a small YIG sphere driven by a microwave field with frequency $\omega_{\rm{d}}$, when its [100] crystallographic axis is aligned parallel to the static magnetic field, the cavity magnonics system has the Hamiltonian (setting $\hbar=1$)~\cite{Wang-16}
\begin{eqnarray}
H&=&\omega_{\rm{c}}a^{\dag}a+\omega_{\rm{m}}b^{\dag}b+Kb^{\dag}bb^{\dag}b+g_{\rm{m}}(a^{\dag}b+ab^{\dag})
 \nonumber\\
&&+\Omega_{\rm{d}}(b^{\dag}e^{-i\omega_{\rm{d}}t}+
be^{i\omega_{\rm{d}}t}),
\label{Hamiltonian}
\end{eqnarray}
where $a^{\dag}(a)$ is the creation (annihilation) operator of the cavity photons at frequency $\omega_{\rm{c}}$, $b^{\dag}(b)$ is the creation (annihilation) operator of the Kittel-mode magnons at frequency $\omega_{\rm{m}}$, and $\Omega_{\rm{d}}$ is the drive-field strength. As shown in \cite{Wang-16}, the Kerr term $Kb^{\dag}bb^{\dag}b$, with a positive coefficient $K=\mu_0K_{\mathrm {an}}\gamma^2/(M^2V_m)$, is intrinsically due to the magnetocrystalline anisotropy in the YIG material. Here $\mu_0$ is the magnetic permeability of free space, $K_{\mathrm {an}}$ is the first-order anisotropy constant, $\gamma=g\mu_B/\hbar$ is the gyromagnetic ratio (with $g$ being the $g$-factor and $\mu_B$ the Bohr magneton), $M$ is the saturation magnetization, and $V_m$ is the volume of the YIG sphere. Note that the Kerr effect is strengthened when reducing $V_m$.

We consider the case with $|\omega_{\rm{LP}}-\omega_{\rm{d}}|\ll|\omega_{\rm{UP}}-\omega_{\rm{d}}|$, where $\omega_{\rm{UP}}$ is angular frequency of the {\it upper}-branch CMPs. In such a case, the drive field applied to the YIG sphere generates polaritons in the lower branch much more than the polaritons in the upper branch. Using a quantum Langevin approach, we obtain a cubic equation for the CMP frequency shift $\Delta_{\rm{LP}}$~\cite{supplementary}
\begin{equation}
\bigg[(\Delta_{\rm{LP}}+\delta_{\rm{LP}})^2+\bigg(\frac{\gamma_{\rm{LP}}}{2}\bigg)^{2}\bigg]\Delta_{\rm{LP}}-cP_{\rm{d}}=0,
\label{cubic}
\end{equation}
where $\gamma_{\rm{LP}}$ is the damping rate of the {\it lower}-branch CMP and $c$ is a coefficient characterizing the coupling strength between the drive field and the {\it lower}-branch CMPs. This cubic equation provides a steady-state solution for the frequency shift of the {\it lower}-branch CMPs as a function of both the drive-field frequency detuning $\delta_{\rm{LP}}$ and the drive power $P_{\rm d}$. Under appropriate conditions, it has three solutions, two of them stable and the additional one unstable. This corresponds to the bistability of the system. In Fig.~\ref{fig:2}, we also show the theoretical results (dashed curves) obtained using Eq.~(\ref{cubic}), which fit the experimental results very well. This verifies the experimental observation of the CMP bistability in our cavity magnonics system. The two stable solutions of $\Delta_{\rm{LP}}$ in Eq.~(\ref{cubic}) correspond to {\it two} states of the system with {\it large} and {\it small} number of polaritons in the lower branch~\cite{supplementary}. Thus, in Fig.~\ref{fig:2}, each sharp switching of the frequency shift $\Delta_{\rm{LP}}$ is related to the transition between these two states.

\begin{figure}
\includegraphics[width=0.48\textwidth]{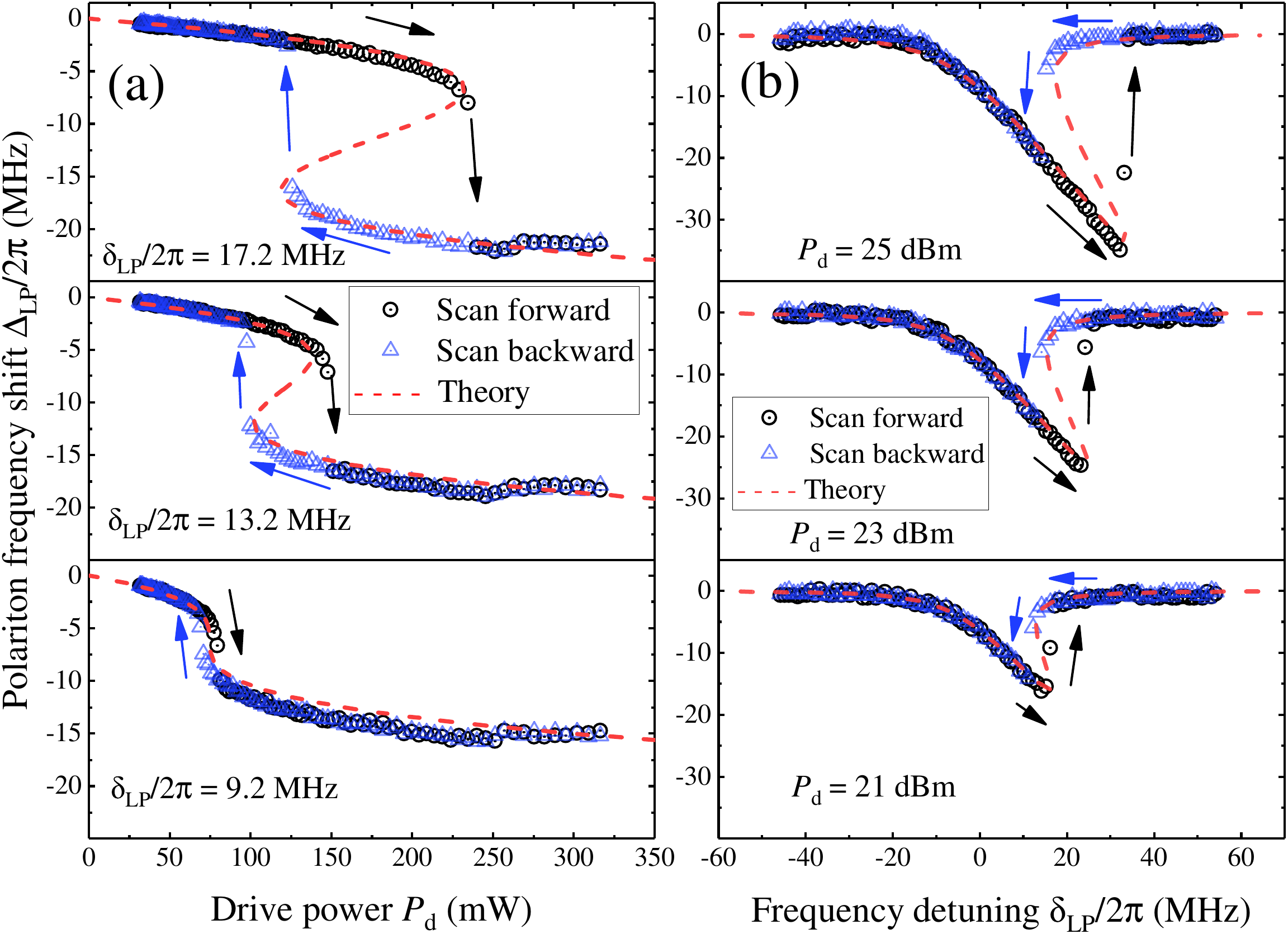}
\caption{(a)~$\Delta_{\rm LP}/2\pi$ versus $P_{\rm d}$ when $\delta_{\rm{LP}}/2\pi=17.2$, $13.2$, and $9.2$~MHz. (b)~$\Delta_{\rm LP}/2\pi$ versus $\delta_{\rm{LP}}/2\pi$ when $P_{\rm d}=25$, 23, and 21~dBm.
The black circle (blue triangle) dots are the forward (backward)-scanning results.
The red dashed curves are theoretical results obtained using Eq.~(\ref{cubic}), with $\gamma_{\rm{LP}}/2\pi=10.65$~MHz. In (a), $c/(2\pi)^3$ is fitted to be $-4$, $-3.5$, and $-3.1$~MHz$^{3}$/mW for curves from top to bottom. In (b), $c/(2\pi)^3$ is fitted to be $-3.01$, $-3.46$, and $-3.6$~MHz$^{3}$/mW for curves from top to bottom. The [110] crystallographic axis of the YIG sphere is aligned parallel to $B_0$ and the magnons are in resonance with $\rm{TE}_{\rm{102}}$.}
\label{fig:3}
\end{figure}

In the resonance case with $\omega_{\rm m}=\omega_{\rm c}$, we further implement experiment by aligning the [110] crystallographic axis of the YIG sphere parallel to the static magnetic field $B_0$.
In Fig.~\ref{fig:3}(a), we measure the frequency shift $\Delta_{\rm LP}$ of the {\it lower}-branch CMPs versus the drive power $P_{\rm d}$ for different values of the drive-field frequency detuning $\delta_{\rm{LP}}$ relative to the {\it lower}-branch CMPs. In sharp contrast to Fig.~\ref{fig:2}(a), bistability of the {\it lower}-branch CMPs is now observed at $\delta_{\rm{LP}}>0$. The area of the hysteresis loop also decreases when reducing $\delta_{\rm{LP}}$, but the hysteresis loop becomes clockwise. Also, we present in Fig.~\ref{fig:3}(b) the frequency shift $\Delta_{\rm LP}$ versus the frequency detuning $\delta_{\rm{LP}}$ for different values of $P_{\rm d}$. The hysteresis loop is also counter-clockwise, similar to that in Fig.~\ref{fig:2}(b).

As shown in \cite{supplementary}, when the YIG sphere is aligned with its [110] crystallographic axis parallel to the static magnetic field $B_0$, the Hamiltonian of the cavity magnonics system takes the same form as in Eq.~(\ref{Hamiltonian}), but the coefficient of the Kerr term becomes negative, $K=-13\mu_0K_{\mathrm {an}}\gamma^2/(16M^2V_m)$.
The corresponding theoretical results (dashed curves) obtained using Eq.~(\ref{cubic}) are shown in Fig.~\ref{fig:3}, which are also in good agreement with the experimental results.
Moreover, Figs.~\ref{fig:2} and \ref{fig:3} show that the CMPs have blue(red)-shift in frequency when the [100] ([110]) crystallographic axis of the YIG sphere is aligned along the direction of the static magnetic field. These are due to the positive and negative Kerr coefficients $K$'s in the two different cases~\cite{supplementary}.

\begin{figure}
\centering
\includegraphics[width=0.48\textwidth]{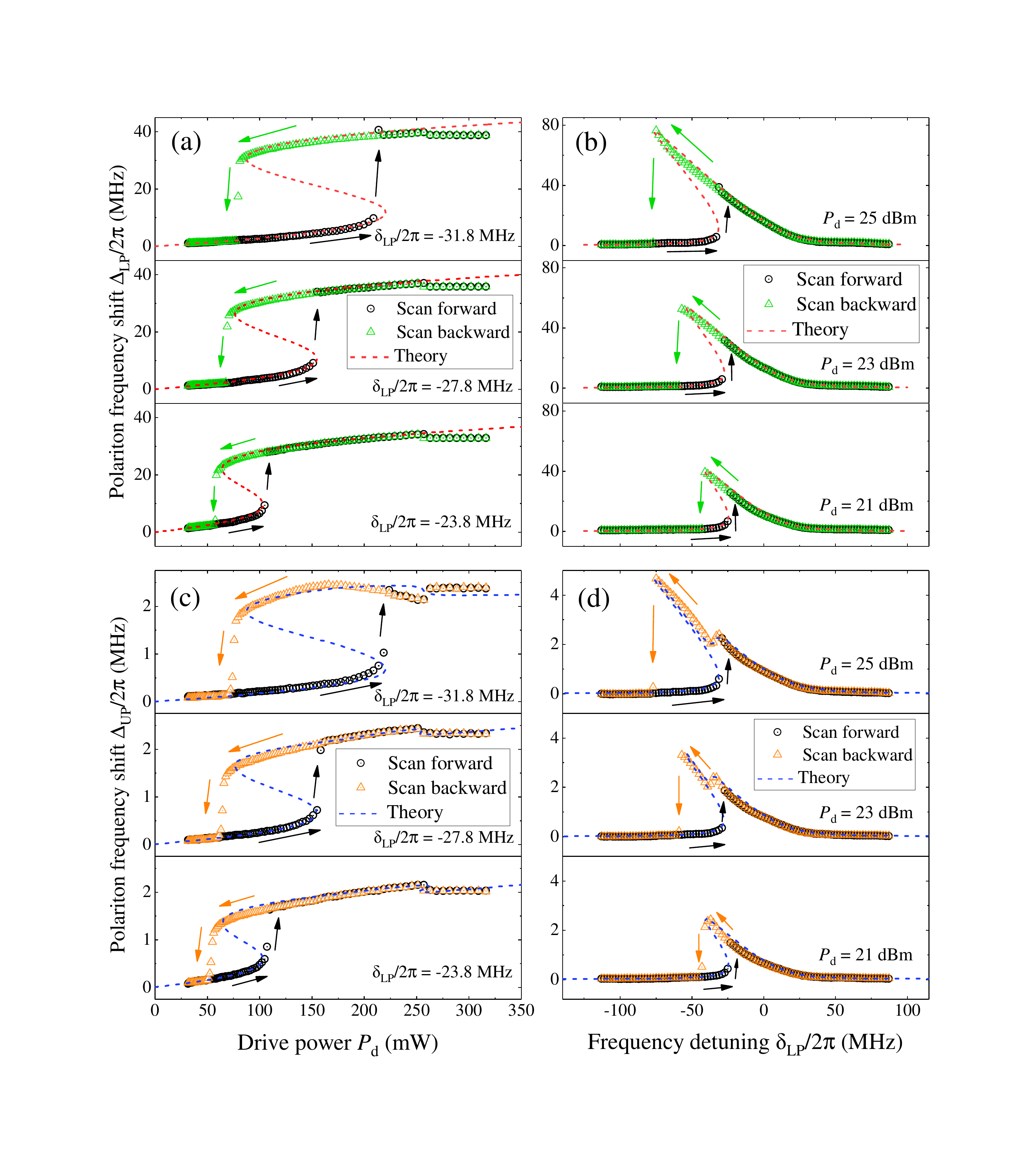}
\caption{(a)~$\Delta_{\rm LP}/2\pi$ versus $P_{\rm d}$ when $\delta_{\rm{LP}}/2\pi=-31.8$, $-27.8$, and $-23.8$~MHz. (b)~$\Delta_{\rm LP}/2\pi$ versus $\delta_{\rm{LP}}/2\pi$ when $P_{\rm d}=25$, 23, and 21~dBm.
The black circle (green triangle) dots are the forward (backward)-scanning results.
The red dashed curves are theoretical results obtained using Eq.~(\ref{cubic}), with $\gamma_{\rm{LP}}/2\pi=16.8$~MHz. In (a), $c/(2\pi)^3$ is fitted to be $25.2$, $25.0$, and $25.3$~MHz$^{3}$/mW for curves from top to bottom. In (b), $c/(2\pi)^3$ is fitted to be $17.0$, $19.0$, and $22.7$~MHz$^{3}$/mW for curves from top to bottom.
(c) and (d)~$\Delta_{\rm UP}/2\pi$ measured under the same conditions as in (a) and (b), respectively.
The black circle (oragne triangle) dots are the forward (backward)-scanning results. The blue dashed curves (except for the parts around small dips) are also obtained using Eq.~(\ref{cubic}), with the fitted ratio $\xi\equiv\Delta_{\rm{UP}}/\Delta_{\rm{LP}}$ for curves from top to bottom being $0.060$, $0.062$, and $0.061$ in (c) and $0.062$, $0.063$, and $0.061$ in (d).
The [100] crystallographic axis of the YIG sphere is aligned parallel to $B_0$ and the magnons are far off resonance with $\rm{TE}_{\rm{102}}$.
In addition, the small dips in, e.g., (c) and (d) can be fitted by introducing a magnetostatic mode with the negative Kerr coefficient~\cite{supplementary}.}
\label{fig:4}
\end{figure}

Finally, we demonstrate the Kerr effect of magnons at an off-resonance point much away from $\omega_{\rm m}=\omega_{\rm c}$, where the magnet coil current is $4.8$~A [see the left dashed line in Fig.~\ref{fig:1}(b)]. At this point, the frequency of the {\it lower}-branch CMPs is about $9.91$~GHz and the frequency of the {\it upper}-branch CMPs is about $10.08$~GHz, as indicated by points $B$ and $C$ in Fig.~\ref{fig:1}(b), respectively. Also, the [100] crystallographic axis of the YIG sphere is aligned parallel to the static magnetic field $B_0$.
In Fig.~\ref{fig:4}(a), we present the frequency shift $\Delta_{\rm{LP}}$ of the {\it lower}-branch CMPs versus the drive power $P_{\rm d}$ for different values of the drive-field frequency detuning $\delta_{\rm{LP}}$ relative to the {\it lower}-branch CMPs. Moreover, the frequency shift $\Delta_{\rm{LP}}$ versus the frequency detuning $\delta_{\rm{LP}}$ is shown in Fig.~\ref{fig:4}(b) for different values of $P_{\rm d}$. As in Fig.~\ref{fig:2}, we also see counter-clockwise hysteresis loops. Because the magnon is a dominating component of the {\it lower}-branch CMP at this very off-resonance point, nearly the bistability of magnons is actually observed here, directly owing to the magnon Kerr effect in the YIG sphere. The theoretical results (dashed curves) obtained using Eq.~(\ref{cubic}) also agree well with the experimental results.

Under the conditions same as in Figs.~\ref{fig:4}(a) and \ref{fig:4}(b), we further show the frequency shift $\Delta_{\rm{UP}}$ of the {\it upper}-branch CMPs versus the drive power $P_{\rm d}$ [Fig.~\ref{fig:4}(c)], as well as the frequency shift $\Delta_{\rm{UP}}$ of the {\it upper}-branch CMPs versus the drive-field frequency detuning $\delta_{\rm{LP}}$ relative to the {\it lower}-branch CMPs [Fig.~\ref{fig:4}(d)]. Here, as in Figs.~\ref{fig:4}(a) and \ref{fig:4}(b), the drive field is still applied on the lower branch. Now the cavity photon is a dominating component of the {\it upper}-branch CMP at this very off-resonance point, so it is nearly the bistability of cavity photons that is observed in Figs.~\ref{fig:4}(c) and \ref{fig:4}(d).
As one knows, both optical~\cite{Baas-04,Gibbs-12} and magnetic~\cite{Hu-09} bistabilities are interesting phenomena of nonlinear systems. Here we observe the {\it simultaneous} bistability of both magnons and cavity photons at a very off-resonance point by applying the drive field only on the lower branch, where the optical bistability is achieved via the magnetic bistability.
Also, the experimental results in Figs.~\ref{fig:4}(c) and \ref{fig:4}(d) fit well with the theoretical results (dashed curves), where the fitted ratio $\xi\equiv\Delta_{\rm{UP}}/\Delta_{\rm{LP}}$ is close to the theoretical value $0.065$~\cite{supplementary}. In addition, some small dips are observed in, e.g., Figs.~\ref{fig:4}(c) and \ref{fig:4}(d). By fitting with these dips, we attribute them to other polaritons stemming from the coupling between cavity photons and a magnetostatic mode with the negative Kerr coefficient~\cite{supplementary}.

In conclusion, we have demonstrated the bistable behaviors of the CMPs using a hybrid system consisting of microwave cavity photons strongly interacting with the magnons in a YIG sphere.
We find that the switching points where the CMPs transition from one state to another depend on the drive power and the drive-field frequency detuning.
When implementing the experiment, we align, respectively, the [100] and [110] crystallographic axes of the YIG sphere parallel to the static magnetic field and find very different bistable behaviors in these two cases. We theoretically fit the experimental results well and put these different bistable behaviors down to the Kerr nonlinearity with either a positive or negative coefficient.

The cavity magnonics system possesses the merits of tunability and compatibility with other quantum systems, in which the magnons can be tuned by an external magnetic field and the magnon-photon coupling can be tuned by moving the YIG sphere inside the cavity. With regard to the bistability of the CMPs, the area of the hysteresis loop can be easily controlled by a lower drive power and a tunable drive-field frequency.
This may bring potential applications in realizing low-energy switching devices. Also,
we observe simultaneous bistability of both magnons and cavity photons by tuning the magnetic field.
It paves a new path to control the conversion between magnetic and optical bistabilities. Moreover, our study provides possible routes to explore other nonlinear effects of the system such as the creation of a frequency comb for frequency conversion~\cite{Tobar-12} and the chaos of CMPs. With the available CMP bistability, the cavity magnonics system can also provide a new platform to understand the dissipative phase transition and the related critical phenomena~\cite{Kessler-12,Carr-13,Rodriguez-17}. These open up different directions for future studies.

\begin{acknowledgments}
This work was supported by the National Key Research and Development Program of
China (Grant No.~2016YFA0301200), the NSFC (grant No.~11774022), the MOST 973 Program of China (Grant No.~2014CB848700), and the NSAF (Grant No.~U1330201 and No.~U1530401).
C.M.H. was supported by the NSFC (Grant No.~11429401).
Y.-P. W and G.-Q. Z contribute equally to this work.
\end{acknowledgments}

\appendix

\end{document}